Imaging Grains and Grain Boundaries in Single-Layer Graphene: An Atomic Patchwork Quilt

Pinshane Y. Huang[1]*, Carlos S. Ruiz-Vargas[1]*, Arend M. van der Zande[2]*, William S. Whitney[2], Shivank Garg[3], Jonathan S. Alden[1], Caleb J. Hustedt[4], Ye Zhu[1], Jiwoong Park[3,5], Paul L. McEuen[2,5], David A. Muller[1,5]

1. School of Applied and Engineering Physics, Cornell University, Ithaca, NY 14853, USA

2. Department of Physics, Cornell University, Ithaca, NY 14853, USA

3. Department of Chemistry and Chemical Biology, Cornell University, Ithaca, NY 14853, USA

4. Department of Physics and Astronomy, Brigham Young University, Provo, UT 84602, USA

5. Kavli Institute at Cornell for Nanoscale Science, Ithaca, NY14853, USA

* These authors contributed equally to this work.

**The properties of polycrystalline materials are often dominated by the size of their grains and by the atomic structure of their grain boundaries. These effects should be especially pronounced in two-dimensional materials, where even a line defect can divide and disrupt a crystal. These issues take on practical significance in graphene, a hexagonal two-dimensional crystal of carbon atoms; Single-atom-thick graphene sheets can now be produced by chemical vapor deposition (CVD)[1,2] on up to meter scales[3], making their polycrystallinity almost unavoidable. Theoretically, graphene grain boundaries are predicted to have distinct electronic[4-7], magnetic[8], chemical[9], and mechanical[6,10-12] properties which strongly depend on their atomic arrangement. Yet, because of the five-order-of-magnitude size difference between grains and the atoms at grain boundaries, few experiments have fully explored the graphene grain structure. Here, we use a combination of old and new transmission electron microscope (TEM) techniques to bridge these length**

**scales. Using atomic-resolution imaging, we determine the location and identity of every atom at a grain boundary and find that different grains stitch together predominantly via pentagon-heptagon pairs. Rather than individually imaging the several billion atoms in each grain, we use diffraction-filtered imaging[13] to rapidly map the location, orientation, and shape of several hundred grains and boundaries, where only a handful have been previously reported[14-17]. The resulting images reveal an unexpectedly small and intricate patchwork of grains connected by tilt boundaries. By correlating grain imaging with scanned probe measurements, we show that these grain boundaries dramatically weaken the mechanical strength of graphene membranes, but do not measurably alter their electrical properties. These techniques open a new window for studies on the structure, properties, and control of grains and grain boundaries in graphene and other two-dimensional materials.**

Figure 1a shows a large array of the suspended, single-layer graphene membranes used in this study. We grew predominately single-layer graphene films on copper foils via CVD[1] and transferred them onto holey silicon nitride TEM grids using two different techniques (See Methods, Supplementary Information). One key innovation over previous graphene TEM sample fabrication[18] was gently transferring the graphene onto a TEM grid using a minimum of polymer support and baking the samples in air to remove the polymer without liquid solvents. This produces large arrays of free-standing graphene sheets covering up to 90% of TEM grid holes.

To characterize these membranes at the atomic-scale, we used aberration-corrected annular dark-field scanning transmission electron microscopy (ADF-STEM), where a 60 keV angstrom-scale electron beam is scanned over the sample while the medium- to high-angle scattered electrons are collected. Properly calibrated, this technique images the location and atomic number[19] of each atom and, along with TEM, has been used to study the lattice and atomic defects of graphene and boron nitrene[19-21]. Figure 1b shows an ADF-STEM image of the crystal lattice within a single graphene grain. Away from the grain boundaries, such regions are defect-free.

In Figure 1c, two graphene grains meet with a relative misorientation of 27°, forming a tilt boundary. As highlighted in Figure 1d, the two crystals are stitched together by a series of pentagons, heptagons, and distorted hexagons. The grain boundary is not straight, and the defects along the boundary are not periodic. While the boundary dislocation resembles structures proposed theoretically[10,12], its aperiodicity contrasts with many of these models and will strongly affect the predicted properties of grain boundaries. By analyzing atomic scattering intensities[19], we confirm the boundary is composed entirely of carbon and stable under the 60 keV electron beam. Thus, the polycrystalline graphene is a strongly-bonded, continuous carbon membrane. We also note that many grain boundaries are decorated by lines of surface particles and adsorbates (Supplementary Figure), suggesting that as predicted[9], they may be more chemically reactive than the pristine graphene lattice.

Both (S)TEM, which determines the positions and identities of atomic nuclei, and complementary STM probing valence wavefunctions[14-16,22] are invaluable for understanding the local properties of grain boundaries. Using these atomic-resolution approaches, however, tens to hundreds of billions of pixels would be needed to fully image even a single micron-scale grain, with estimated acquisition times of a day or more. Other candidates for characterizing grains on larger scales, such as LEEM[17] and Raman microscopy[23], typically cannot resolve small grains and may be difficult to interpret. Fortunately, electron microscopy offers an ideal technique to image grains on the necessary length scales: dark-field TEM (DF-TEM), a high-throughput, diffraction-sensitive imaging technique[13] that can be implemented on most TEMs built in the last sixty years. While this method is usually applied to hundred-nanometer-thick foils[13], we demonstrate below that remarkably, it also works on single-atom thick sheets—even on samples too dirty for atomic resolution imaging. In this manner, DF-TEM provides a nanometer- to micron-scale grain analysis that complements ADF-STEM to give a complete understanding of graphene grains on every relevant length scale.

Figures 2a and 2b show a bright-field TEM image of a graphene sheet along with the selected area electron diffraction pattern created from this region of the membrane. Due to graphene's sixfold symmetry, electron diffraction from a single graphene crystal results in one set of sixfold-symmetric spots. Figure 2b contains many such families of spots, indicating that the field of view contains several grains of different orientations. DF-TEM images these grains one-by-one with few-nanometer resolution using an objective aperture filter in the back focal plane to collect electrons diffracted to a small range of angles, as shown by the circle in Figure 2b. The resulting real-space image (Figure 2c) shows only the grains corresponding to these selected in-plane lattice orientations and requires only a few seconds to acquire. By repeating this process using several different aperture filters, then coloring and overlaying these DF-images, (Figure 2d,e), we create complete maps of the graphene grain structure, color-coded by lattice orientation, as shown in Figure 2e-g.

The images obtained are striking. The grains have complex shapes and many different crystal orientations. In Figure 2e-g, we observe special locations from which many grains emanate. Small particles and multilayer graphene also are often found near these sites (e.g. Figure 2e, top right). Both the average spacing (2-4 µm) and shapes of these radiant sites are comparable with previous studies of graphene nucleation on copper foils using Raman and SEM[1,23], suggesting that these locations are likely nucleation sites. Similar structures have been observed in studies of crystallization in colloids and are consistent with crystallization around impurities[24]. Significantly, each apparent nucleation site gives rise to many grains of different orientation, resulting in a mean grain size much smaller than the nucleation density.

The distributions of grain size and relative angular orientation are readily determined from DF-TEM images. In Figure 3a, we plot a histogram of grain sizes across several samples. The mean grain size, defined as the square root of the grain area, is $250 \pm 11$ (s.e.m.) nm. This size is much smaller than the grain size of the copper substrate (100 μm- 1mm)[1] and typical lateral grains measured in bulk HOPG (6-30 μm)[25]. We find similar results in samples fabricated from several different growth runs and in two separate CVD furnaces. The inset in Figure 3a shows the cumulative probability of finding multiple grains in a given area. This plot demonstrates that micron-scale CVD graphene devices will nearly always contain multiple grains. Figure 3b shows a histogram of the relative crystallographic angles between adjacent grains. With graphene's sixfold crystal symmetry, the diffractive-imaging technique only determines grain rotations modulo 60°. Consequently, the measurable difference between grain orientations is from 0-30° (i.e. 31° is measured as 29°). We observe a surprising and robust preference for low-angle (~7°)grain boundaries and high (~30°) angle boundaries similar to the one seen in Figure 1.

Additional information about these orientations comes from the larger-area diffraction patterns in Figure 3c, created by averaging diffraction data sampled across 1200 μm$^2$ regions of graphene. The broadened diffraction peaks in Figure 3c (left) show a distinct sixfold pattern, indicating that a significant fraction of the grains are approximately aligned across large areas This alignment can also be seen in Figure 3d, a low-magnification DF-TEM image displaying grains with a small (~10°) range of in-plane lattice orientations. Almost half of the membrane appears bright, indicating these grains are all approximately aligned. In contrast, a DF-image of randomly oriented grains would only show roughly one-sixth (10°/60°) of the graphene membrane. In the diffraction pattern of a separately grown sample, (Figure 3c, right), we instead find a clear 12-fold periodicity, indicating that there are two main families of grains rotated 30° from one another. These distributions, which often contain smaller sub-peaks (see Supplementary Figures), are consistent with the frequent observation of low and high ~30° grain boundaries. We attribute these alignments to registry to the copper substrate used for graphene growth. Such registry has recently been observed in STM studies of graphene growth on Cu (100) and (111) surfaces[14,15].

The ability to easily image the grain structure in graphene monolayers opens the door to the systematic exploration of the effects of grain structure on the physical, chemical, optical, and electronic properties of graphene membranes. We find such studies are further facilitated because grain boundaries are visible in scanning electron microscopy and atomic force microscopy (AFM) phase imaging due to preferential decoration of the grain boundaries with surface contamination (See Figure 4a and Supplementary Figures). Below, we show two examples probing the electrical and mechanical properties of grain boundaries.

We first examine the failure strength of these polycrystalline CVD graphene membranes using AFM. We used AFM phase imaging to image grains and then pressed downward with the AFM tip to test the mechanical strength of the membranes. As seen in Figure 4b, the graphene tears along the grain boundaries. From repeated measurements, we find that failure occurs at loads of about ~ 100 nN, an order of magnitude lower than typical fracture loads of 1.7 µN reported for single-crystal exfoliated graphene[26]. Thus, the strength of polycrystalline graphene is dominated by the grain boundaries.

We next probed the electrical properties of these polycrystalline films using suspended membrane devices[27], shown schematically in Figure 4c. Using transport measurements and AC-electrostatic force microscopy (AC-EFM)[28], we find room-temperature mobilities of 800 - 4,000 cm$^2$/V·s, comparable to previous results on CVD graphene[1,27] and only marginally less than mobilities reported for exfoliated graphene (1,000 - 20,000 cm$^2$/V·s)[29]. Figure 4c shows the relative potential along a graphene membrane between two biased electrodes measured using AC-EFM. In this plot, high-resistance grain boundaries would manifest as sharp drops in potential. Using the mean grain size from Figure 3a, a line scan across these 3 μm long membranes should cross an average of 12 grains. However, no noticeable potential drops were detected, indicating that most grain boundaries in these devices are not strongly resistive interfaces. By assuming a grain boundary running perpendicular to the line scan, we estimate an upper bound on the average grain boundary resistance of $R_{GB}$ < 60 Ω-μm/$L$, where $L$ is the length of the grain boundary, compared to $R_{graphene}$ = 700 Ω/□ for the entire device. Measurements on five additional graphene membranes, both suspended and unsuspended, produced similar results. In other words, the resistance of the grain boundaries is less than one-third the resistance of an average-sized grain. This stands in stark contrast to other materials such as complex oxides, where a single grain boundary can lead to a million-fold increase in resistance over single crystals[30].

The imaging techniques reported here provide the tools to characterize graphene grains and grain boundaries on all relevant length scales. These methods will be crucial both for exploring synthesis strategies to optimize grain properties and for studies, such as those demonstrated above, on the microscopic and macroscopic impact of grain structure on graphene membranes. Thus, these results represent a critical step forward in realizing the ultimate promise of atomic membranes in electronic, mechanical and energy-harvesting devices.

**Methods Summary**

**TEM/STEM**

ADF-STEM imaging was conducted using a NION ultraSTEM100 with imaging conditions similar to those used in Krivanek et al[19]. At 60 kV using a 33-35 mrad convergence angle, the electron probe was close to 1.3Å in size and did not damage the graphene. Images presented in the main text were acquired with the medium-angle annular dark field detector with acquisition times between 16 and 32 μs per pixel. For TEM imaging, we used a FEI Technai T-12 operated at 80 kV. Acquisition time for DF images were 5-10 seconds per frame. The spatial resolution in DF-images ranges from 1-10 nm and is set by the size of the objective filtering aperture, in a trade-off between real-space resolution and angular resolution in reciprocal space.

**Scanned probe measurements**

For AFM deflection measurements, we used a MFP3D scope from Asylum Research. We employed silicon AFM probes (Multi75Al, Budget Sensors) with a Resonant Freq. of ~ 75 kHz and force constant of ~3 N/m and tip radius of < 10 nm. All imaging was done in tapping mode. For AC-EFM measurements, a DI 4100 AFM with a Signal Access Module was operated in Lift Mode with $V_{tip}$ = 2 V, a lift height of 10 nm and no piezo drive on the tip. An AC voltage $V_0$ = 1 V was applied through the electrodes at the resonance frequency of the EFM cantilever $f_{cant}$ ~ 77 kHz. An electrostatic force drives the EFM cantilever to resonate, and the amplitude of motion is measured.

**Acknowledgements**

The authors acknowledge discussions with Mark Levendorf, Sharon Gerbode, Houlong Zhuang, Earl Kirkland, John Grazul, Judy Cha, Lena Fitting-Kourkoutis, and Melina Blees. This work was supported in part by the NSF through the Cornell Center for Materials Research and the Nanoscale Science and Engineering Initiative of the National Science Foundation under NSF Award EEC-0117770,064654. Additional support was provided by the Army Research Office, CONACYT-Mexico, the Air Force Office of Scientific Research, and the MARCO Focused Research Center on Materials, Structures, and Devices. Sample Fabrication was performed at the Cornell node of the National Nanofabrication Infrastructure Network, funded by the NSF. Additional facilities support was provided by the Cornell Center for Materials Research (NSF DMR-0520404 and IMR-0417392) and NYSTAR.

**Author Contributions**

P.Y.H., C.S.R, and A.M.V. contributed equally to this work. Electron microscopy and data analysis were carried out by P.Y.H. and D.A.M., with Y.Z. contributing to initial DF-TEM; Graphene growth and sample fabrication by A.M.V and C.S.R under supervision of P.L.M and J.P., aided by S.G, W.S.W., J.S.A, and C.J.H. AC-EFM and analysis by A.M.V. and P.L.M. AFM mechanical testing and analysis by C.S.R. and J.P, aided by S.G. All authors discussed the results and implications at all stages. P.Y.H, A.M.V., C.S.R, P.L.M, J.P., and D.A.M. wrote the paper.

**Figure 1| Atomic-resolution ADF-STEM images of graphene crystals. a,** An SEM image of graphene transferred onto a TEM grid with over 90% coverage via novel high-yield methods. Scale bar 5 μm. **b,** An ADF-STEM image shows the defect-free hexagonal lattice inside a graphene grain. **c,** Two grains (bottom left, top right) intersect with a 27° relative rotation. An aperiodic line of defects stitches the two grains together. **d,** The image from (c) is overlaid with a trace of pentagons (blue), heptagons (red), and distorted hexagons (green). Images (b-d) were low-pass filtered to remove noise; Scale bars in (b-d) are 5 Å.

**Figure 2| Large-scale grain imaging via DF-TEM.** (a-e), Grain imaging process. **a,** Samples appear uniform in bright-field TEM images. **b,** A diffraction pattern taken from a region in (a) reveals that this area is polycrystalline. Placing an aperture in the diffraction plane filters the scattered electrons forming **c,** a corresponding DF-image showing the real-space shape of these grains. **d,** Using several different aperture locations and color-coding them produces **e,** a false-color DF-image overlay depicting the shape and lattice orientation of several grains. **f-g,** Images of regions where many grains emanate from a few points. Scale bars 500 nm.

**Figure 3| Statistical analysis of grain size and orientation. a,** A histogram of grain sizes, taken from three representative samples using DF-TEM. The mean grain size is 250 ± 11 nm (s.e.m., n=535). **a inset,** Plot of the cumulative probability of having more than one grain given the area of a device. **b,** A histogram of relative grain rotation angles measured from 238 grain boundaries. **c,** Large-area diffraction patterns and **d,** a low-magnification DF-TEM image show that grains are globally aligned near particular directions. Scale bar 2 μm.

**Figure 4| AFM indentation and AC-EFM studies of graphene grain boundaries.** (a,b), AFM phase images of a graphene grain before and after an indentation measurement. **a,** Indentation takes place at the center of this grain as shown by the arrow. **b,** The region is torn along grain boundaries after indentation. Scale bars 200 nm. **c,** Electrostatic potential, averaged from three adjacent line scans along a suspended graphene sheet between two electrodes (schematic above) and measured using AC-EFM. Though on average each line scan should cross 12 grains, no measureable features are present. Dashed lines indicate the locations of the electrodes.

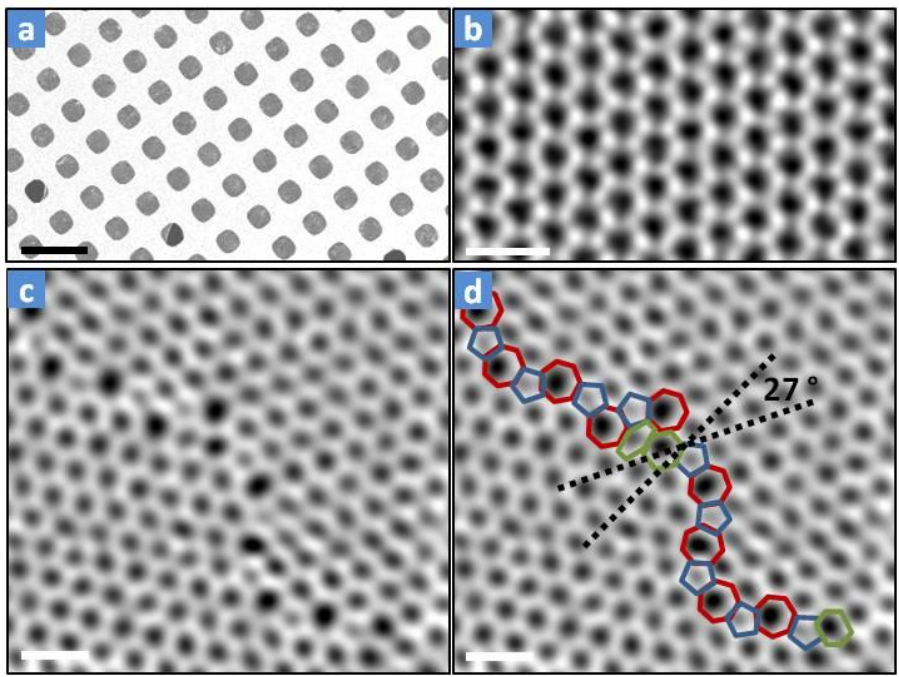

**Figure 1**

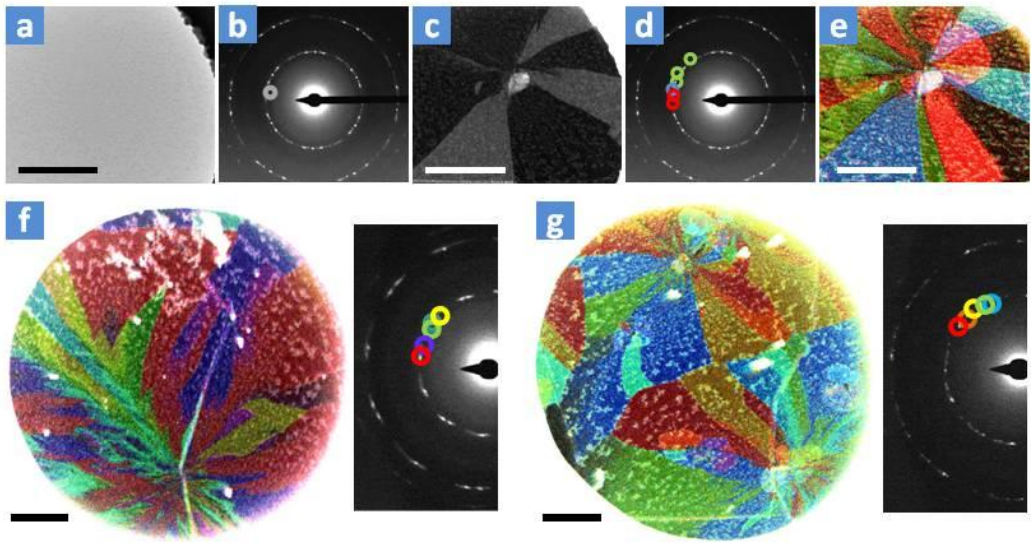

**Figure 2**

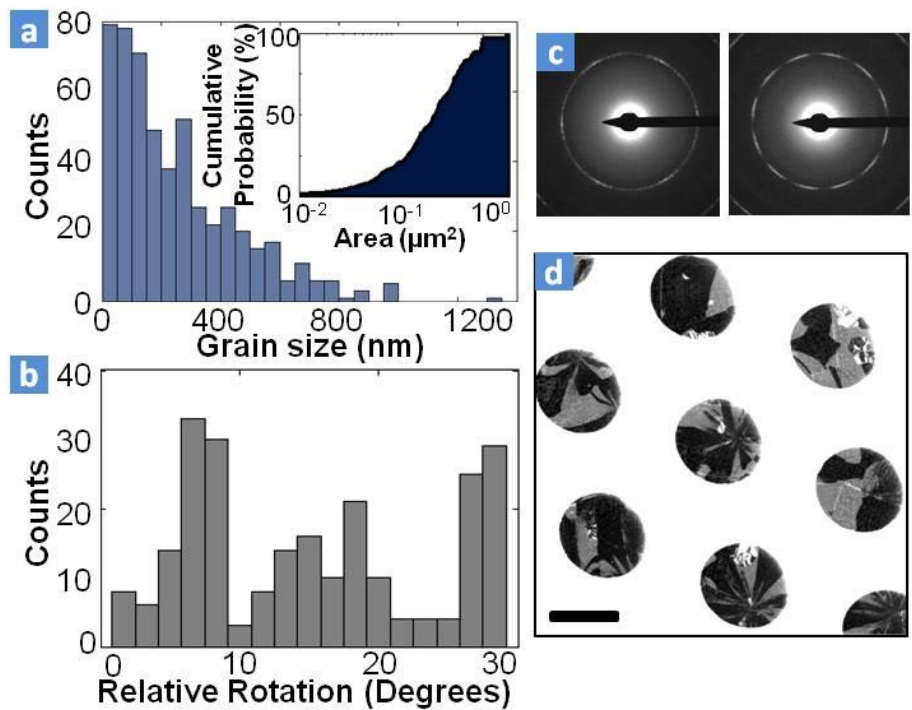

**Figure 3**

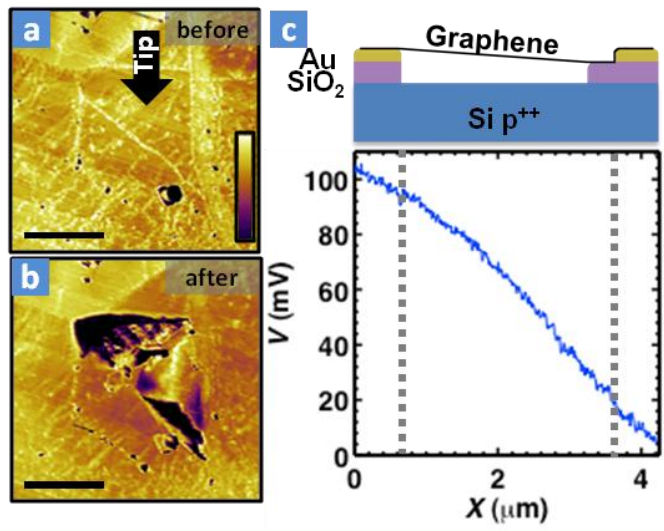

**Figure 4**

## Methods

### ADF-STEM

ADF-STEM imaging was conducted using a NION ultraSTEM 100, operated at 60kV. Imaging conditions were similar to those used in Krivanek et al[18]. Using a 33-35 mrad convergence angle, our probe size was close to 1.3Å. Because the low-voltage electron beam was below the damage threshold energy, the pristine graphene lattice remains stable and defect-free. Images presented in the main text were acquired with the medium-angle annular dark field detector with acquisition times between 16 and 32 µs per pixel.

### DF-TEM

TEM imaging was conducted using a FEI Technai T-12 operated at 80 kV, which did not cause any apparent damage to the graphene membranes. Acquisition time for DF images were 5-10 seconds per frame. The spatial resolution in DF-images ranges from 1-10 nm and is set by the size of the objective filtering aperture in a trade-off between real-space resolution and angular resolution in reciprocal space.

### AC-EFM

A DI 4100 AFM with a Signal Access Module was operated in Lift Mode with $V_{tip}$ = 2 V, a lift height of 10 nm and no piezo drive on the tip. An AC voltage $V_0$ = 1 V is applied through the electrodes at the resonance frequency of the EFM cantilever $f_{cant}$ ~ 77 kHz. An electrostatic force drives the EFM cantilever to resonate, and the amplitude of motion is measured.

### AFM Imaging and Deflection Measurements

For AFM deflection measurements, we used a MFP3D scope from Asylum Research. We employed silicon AFM probes (Multi75Al, Budget Sensors) with a resonant freq. of ~ 75 kHz, a force constant of ~3 N/m, and a tip radius of < 10 nm. All imaging was done in tapping mode. Images were taken with resolutions of 512 x 512, or 1024 x 1024, with acquisition times of at most 10 minutes.

### Sample Fabrication:

We grew single-layer graphene on Cu foils using the chemical vapor deposition method as described by Li et al[1] and then transferred the films to commercial holey SiN TEM grids (such as PELCO Holey Silicon Nitride Support Films) with 2.5 micron-diameter holes for TEM measurements, or larger holes for AFM measurements.

### Samples for DF-TEM

The fabrication for DF-TEM samples is a gentle graphene transfer method using a thin PMMA support, which produced roughly 90% coverage of TEM grid-holes (i.e. 90% of the grid holes, were uniformly covered with suspended graphene, with each TEM grid containing more than 10,000 holes in its nitride window). In this method, after graphene growth on a copper foil, a thin layer of PMMA is spun onto the graphene (2% in anisole, 4000 rpm for 30 seconds), without a post-baking step. Copper is then etched away by floating the foil, PMMA side up, in a HCl/FeCl3 copper etchant (Transene, Type 100/200). Next, the graphene and polymer support is washed by transferring them to DI water baths, taking care to not bring the PMMA in contact with liquids to avoid depositing unwanted residues on the PMMA side of this layer. Finally, the PMMA/graphene layer is scooped out in pieces onto TEM grids. PMMA can be thermally decomposed[31], a gentler process than using liquid solvent rinses. We baked our samples in air (350°C for 3-4 hours), without the use of Ar flow, which can slow the cleaning effect substantially. This step removes the PMMA layer, leaving the graphene freely suspended in a liquid-free release process. These high-yield samples were used in DF-TEM because they provided enough clean graphene to image large numbers of grains.

**Samples for ADF-STEM**
Our secondary technique produced cleaner, but lower-yield graphene using a polymer-free transfer method. This technique is similar to the methods of Regan et al.,[17] in which TEM grids are placed on top of the foil prior to etching and attached by dropping methanol on the grids. Our main addition to this technique was to bake the final samples in a series of anneals increasing in temperature. The grids were then baked in air at 350°C for 2 hours. A final step was to anneal the samples briefly in UHV at 950°C and then 130°C for >8 hours. Because this transfer method uses no support film for the graphene as it is transferred, this method was a comparatively low-yield transfer process with just a few percent coverage over the holes. The advantage to this technique over the polymer-based transfer was that it produced graphene with less surface carbon –areas of hundreds of nanometers appeared atomically clean in ADF-STEM images.

**Electrically Contacted Samples:**
We fabricated electrically contacted suspended graphene by growing single layer graphene, patterning the graphene into 3 μm wide strips while still on the copper foil using contact lithography, and transferring the patterned strips onto a substrate with pre-patterned gold electrodes and trenches.

31      Jiao, L. *et al.* Creation of Nanostructures with Poly(methyl methacrylate)-Mediated Nanotransfer Printing. *Journal of the American Chemical Society* **130**, 12612-12613 (2008).

# Supplementary Figures

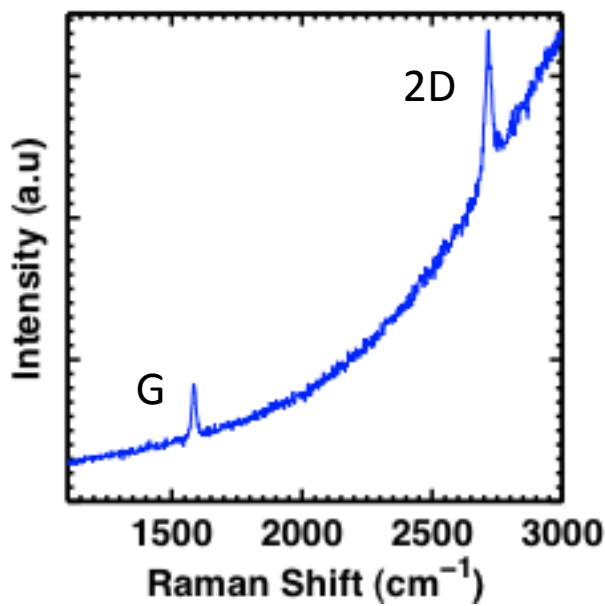

**Supp. Figure 1| Representative Raman spectrum of single-layer graphene samples on copper.** The shape and size of the G and 2D peaks show that the graphene is predominantly single layer. In most samples, we measure only a very small D peak, indicating that we are growing graphene with little disorder. The spectrum is taken before transfer to the TEM grid.

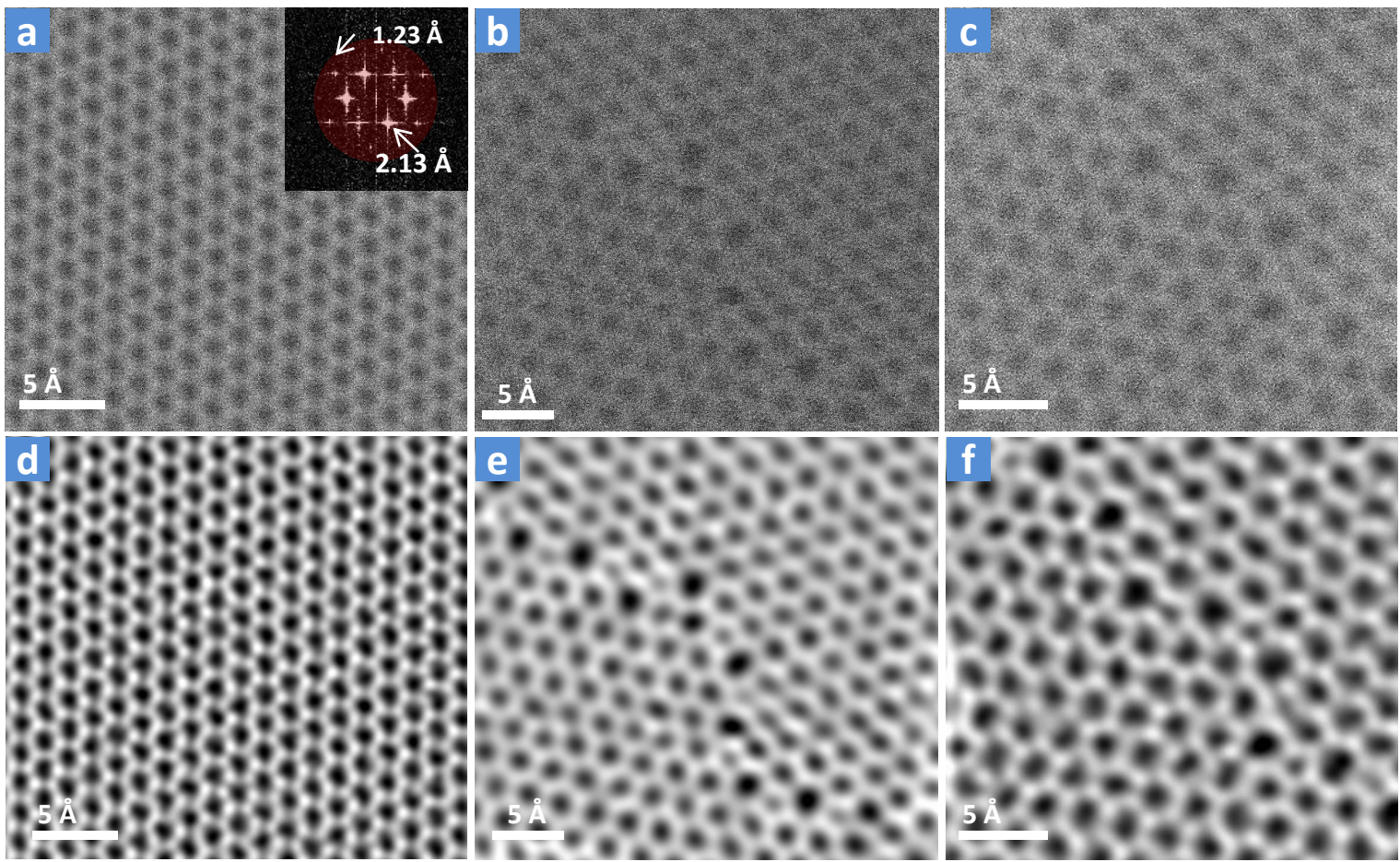

**Supp. Figure 2| Raw and low-pass filtered images of the graphene lattice and grain boundaries. a-c,** Raw ADF-STEM images with Fourier transforms (inset) and masks (insets, color) used to produce the filtered images in **d-f.** The Fourier region inside the masks are applied with a smoothing region of 5-10 pixels and then inverse transformed to form the images in (d-f).

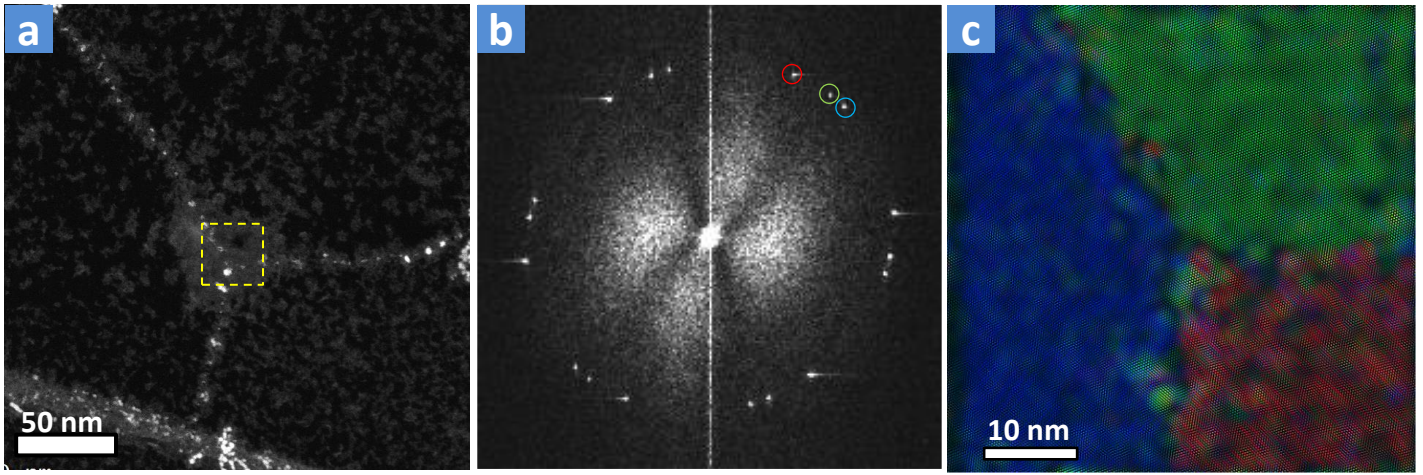

**Supp. Figure 3| Lines of surface contamination mark grain boundaries. a,** Low-magnification ADF-STEM image of a region with three lines of contamination (lines meeting at center) and a wrinkle in the graphene (thick band, lower left) **b,** Fast fourier transform (FFT) of a bright-field STEM image of the region in the area marked in (a). **c,** BF-STEM image formed from taking the inverse FFT of (b) using the colored masks marked. This image shows that the lines of contamination, which are ~5-10 nm wide, occur along grain boundaries.

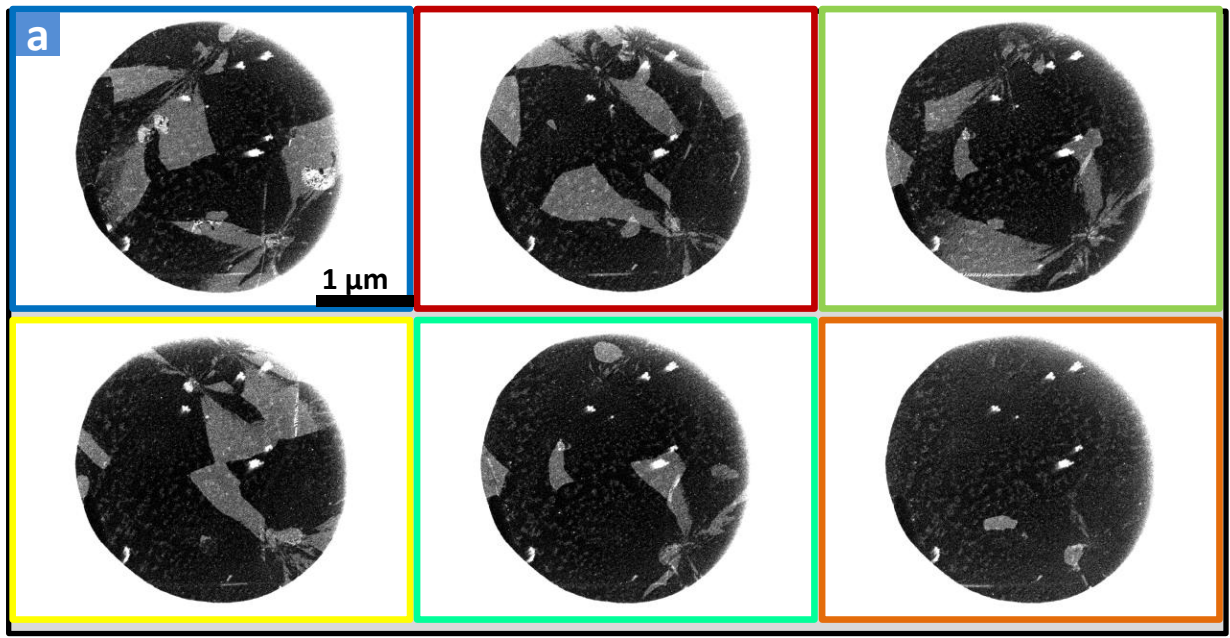
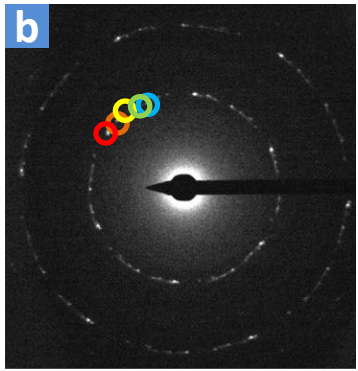
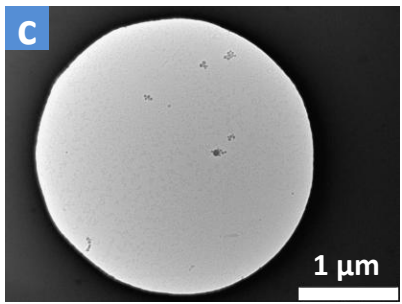
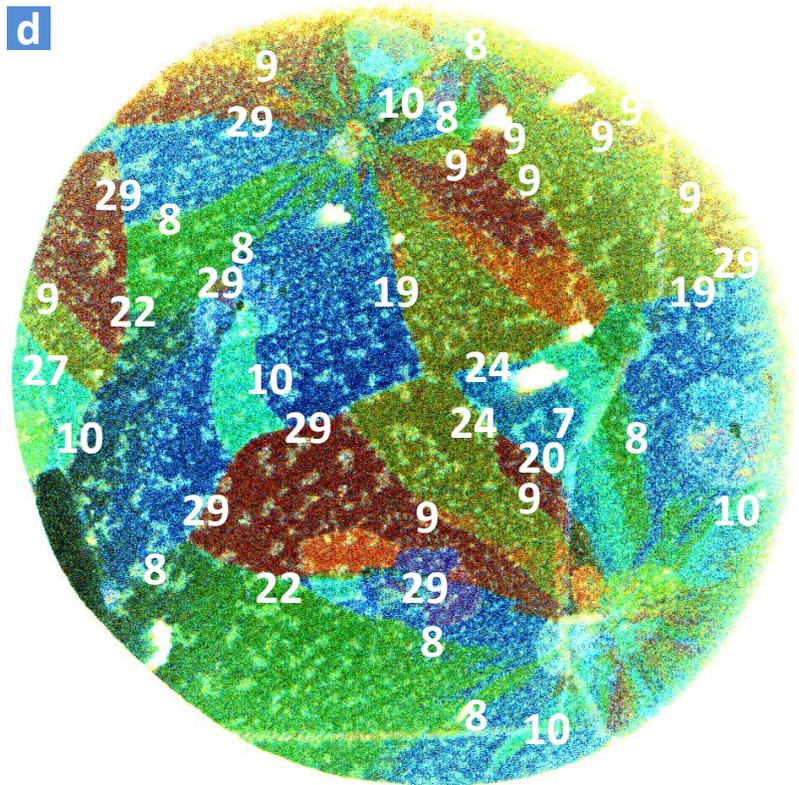

**Supp. Figure 4| DF processing and analysis. a**, Set of raw DF-TEM images for 6 different objective aperture locations, used to create the final composite colored grain image in (d). **b,** Diffraction pattern from a 1 µm diameter region in the grid showing the locations of the apertures, color-coded to match (a) and (d) **c,** bright-field TEM image of the same area, showing no grain contrast. **d,** Final composite image marked with measurements of the grain boundary angles. Most grain boundaries are <10˚ or >20 ˚. Error is ± 2-5 nm.

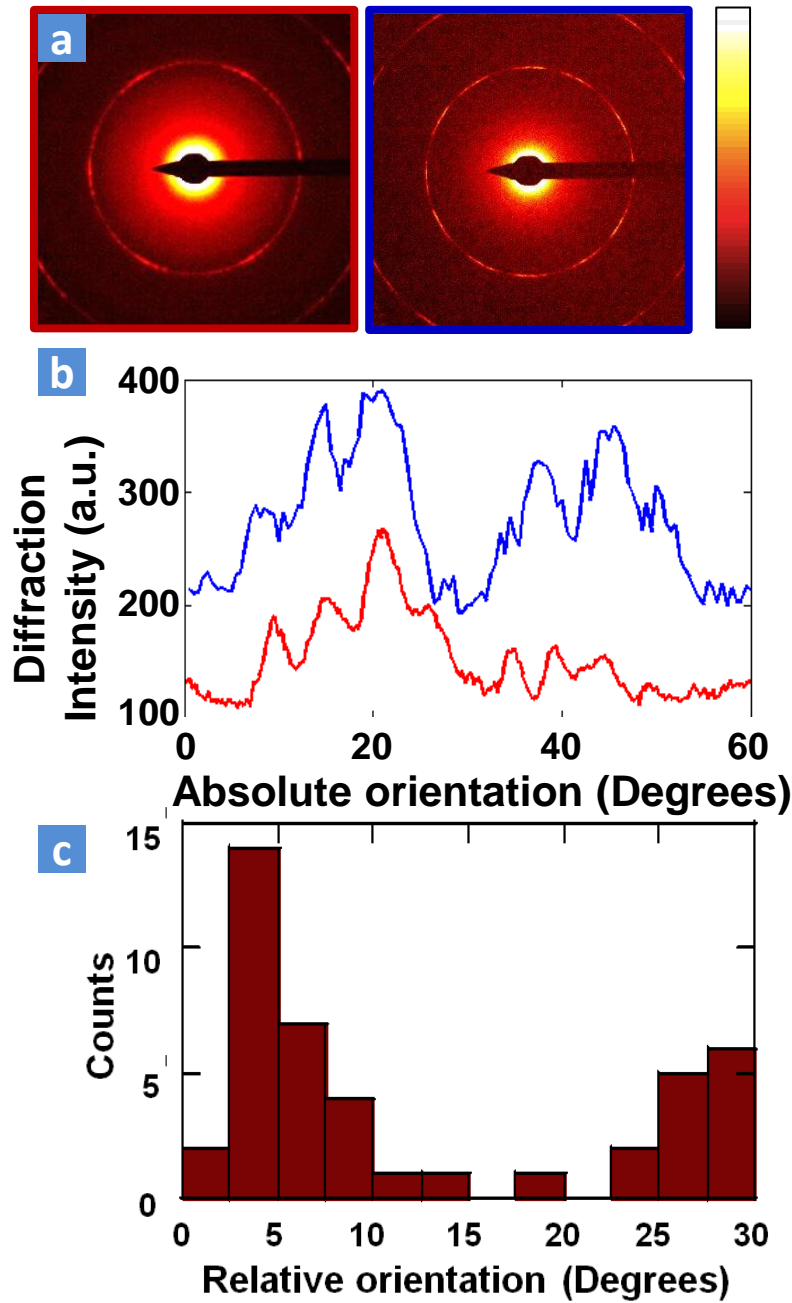

**Supp. Figure 5| Additional data on global and relative grain orientations. a,** Averaged SAED patterns from two samples taken in the same manner as Figure 3c. The left diffraction pattern is from Figure 3c (left). The right pattern is from an additional sample. **b,** Polar plots of the 2.1 Å reflection for the two diffraction patterns in (a). Plots are generated by averaging the diffraction peaks every 60 degree period. **c,** Statistics on relative rotations between grains taken using aberration-corrected ADF-STEM. These statistics are directly comparable to those in Figure 3b.

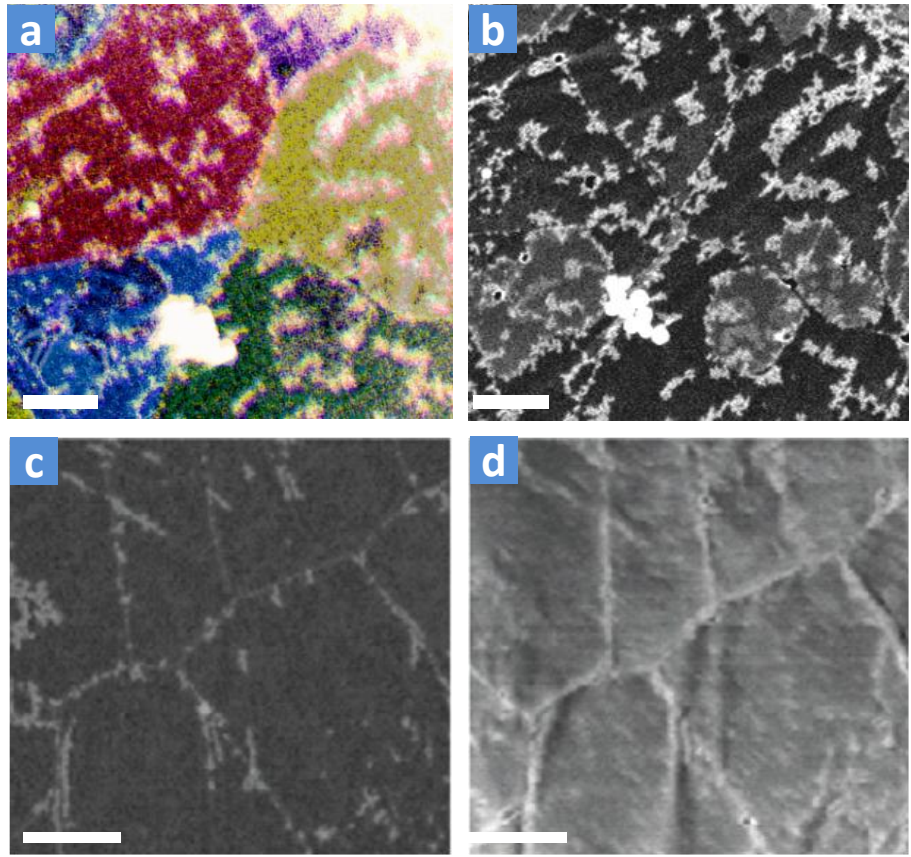

**Supp. Figure 6| Direct comparisons between TEM ,STEM, SEM, and AFM images of grain boundaries. a,** Composite DF-TEM and **b,** SEM images of the same region. We also show similar comparisons between **c,** ADF STEM, and **d,** AFM phase images of a second region. Decorated grain boundaries are visible in SEM, STEM, and AFM phase images. Scale bars are 250 nm.

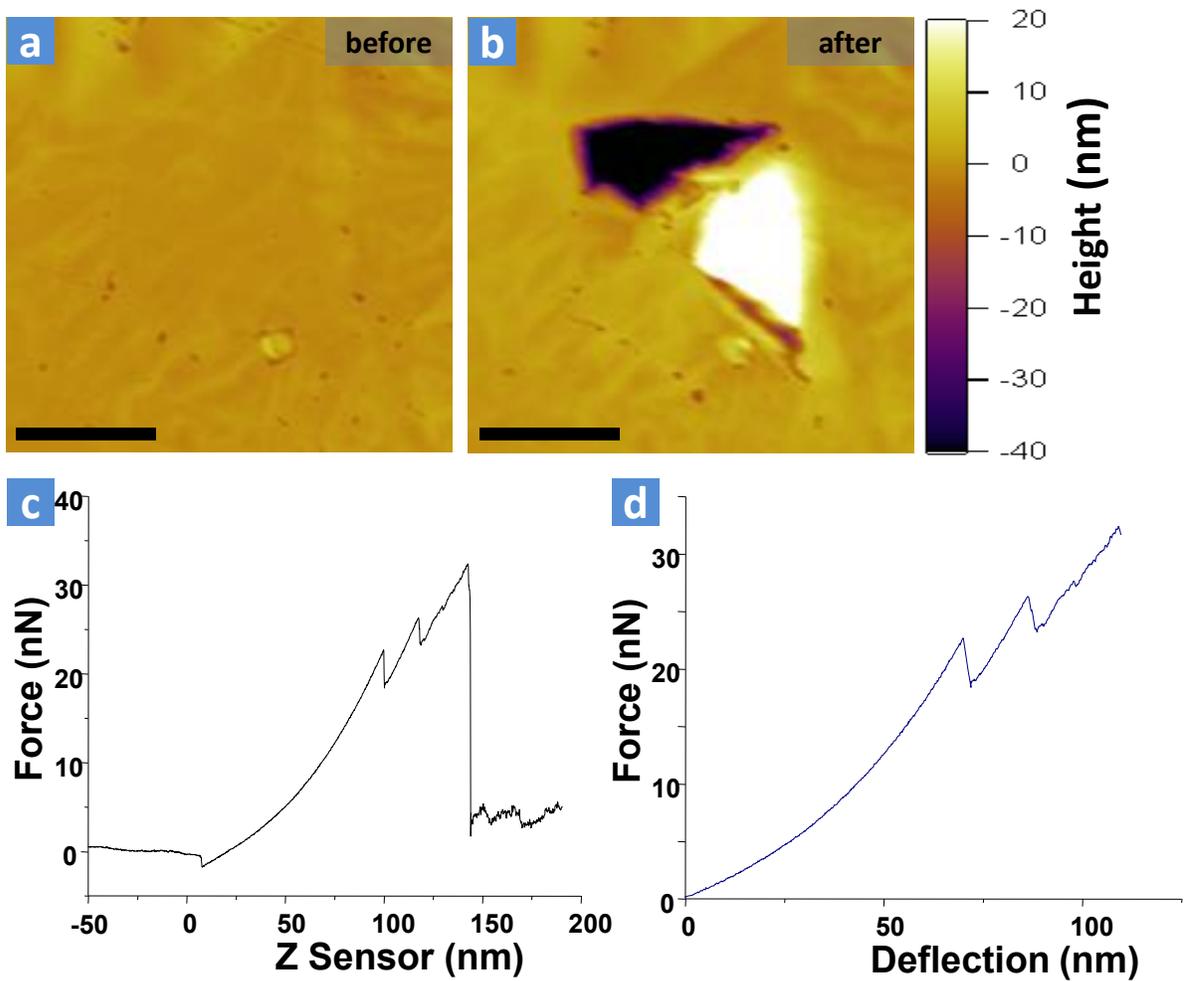

**Supp. Figure 7| AFM indentation curves and topography.** AFM topography images **a,** before and **b,** after the indentation measurement resulting in the tearing of graphene sheet along the grain boundaries of a single domain. **c,** Force plot exerted by the AFM tip as a function of the z position of the AFM piezo. At a relatively low force of about 30 nN, the force drops back down due to the graphene sheet breaking. Before this, smaller drops in the force likely corresponding to smaller tears can be observed. **d,** Corrected force plot, taking into account the tip's deflection, corresponding to the actual graphene's vertical deflection under the AFM tip. Scale bars are 200 nm.

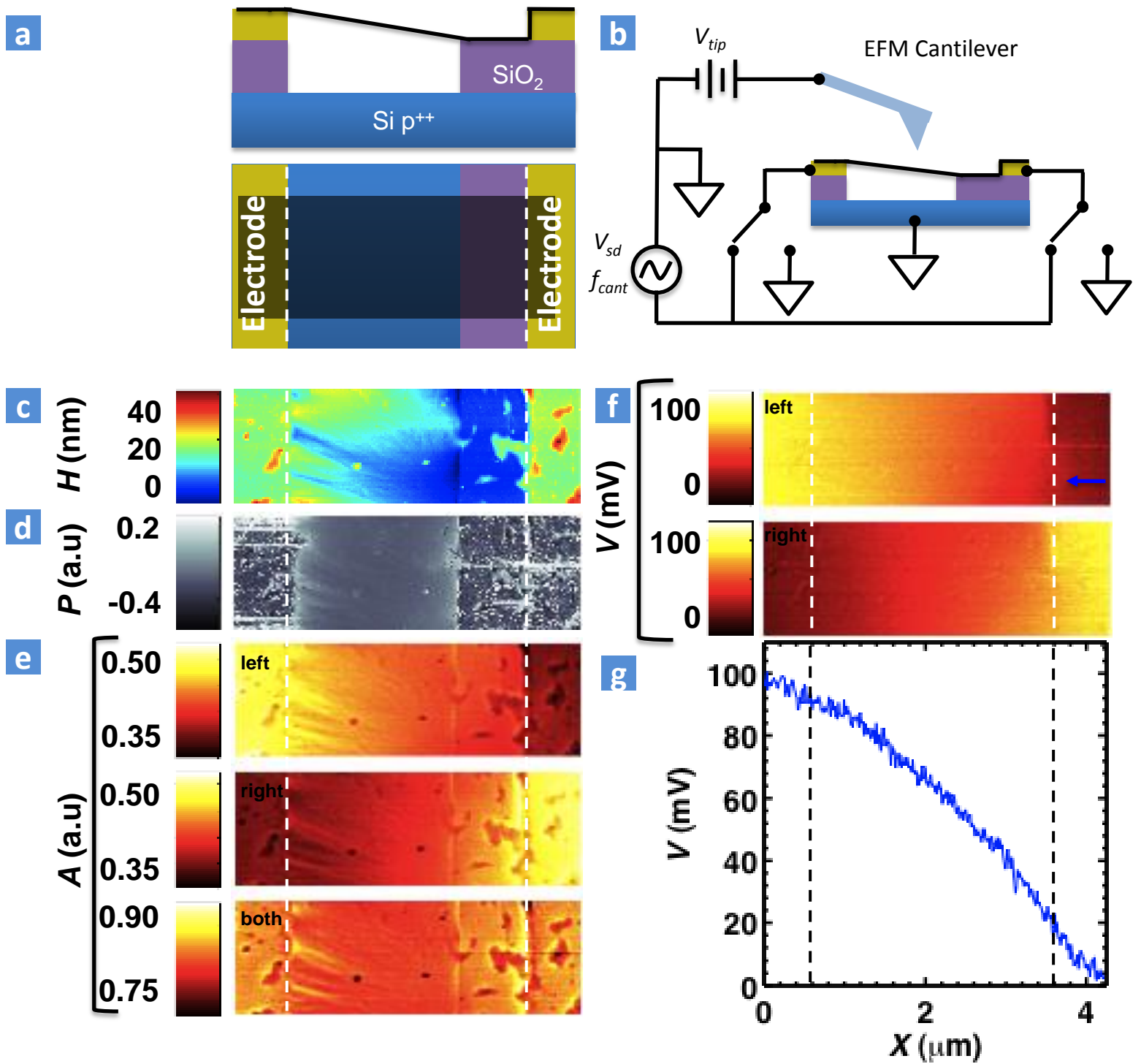

**Supp. Figure 8 | AC-EFM data and processing. a,** Side and top schematics of suspended electrically contacted graphene. **b,** Schematic of AC-EFM measurement setup. **c,** AFM topography and **d,** phase images of a suspended electrically contacted sheet of graphene. **e,** AC-EFM images when driving the left, right and both electrodes respectively. **f,** Ratio of left and right driven electrode EFM images to both electrode driven image. This ratio is proportional to electrostatic potential along the sheet. Features due to changing contaminants and topography of the images disappear. Color bar rescaled to exclude external resistance. **g,** Single line trace from ratio image taken along blue arrow in figure (f). All images are 4.2 μm across, and dashed lines indicate electrode locations.

# Supplementary Methods

**Outline:**
1. Raman Spectroscopy
2. ADF-STEM image processing
3. Low magnification STEM imaging of grain boundaries
4. DF-TEM image overlay procedure
5. DF-TEM statistics acquisition
6. More statistics on grain angles
7. Grain imaging comparisons: DF-TEM, STEM, AFM, SEM
8. AFM indentation
9. AC-EFM

## 1. Raman spectroscopy

Figure S1 shows a representative Raman spectrum of the graphene after growth on copper[1,2]. The shape and relative size of the G and 2D peaks show that the graphene is predominantly single layer. In most samples, we measure only a very small D peak, if it is at all discernable, indicating that we are growing graphene with very little disorder.

## 2. ADF-STEM image processing

Figure S2a-c show the raw STEM images used to produce the images in Figure 1. These images were low-pass filtered to reduce noise using masks similar to the inset in Figure S2a. To create images d and f, the masks were positioned slightly outside the 1.23 Å spot. For Figure S2e, the mask is positioned between the 1.23 and 2.13Å spots—such a tight low-pass filter was employed to make viewing the image and identifying the polygons in the lattice easier. Figure S2d-f are the final images. Figure S2d, in addition to the processing described above, was created by cross-correlating and summing ten lattice images taken in succession on the same region of the graphene lattice to improve signal-to-noise. Finally, Figure S2c/f show that we are able to get atomic resolution images of the grain boundaries.

## 3. Low Magnification STEM imaging of grain boundaries

Figure S3a shows a low magnification STEM image of three grain boundaries intersecting at a point. In this image, grain boundaries appear as bright lines due to adsorbed contamination decorating the grain boundaries. The contamination lines are roughly 4-15 nm wide in STEM. In addition, there is a much wider (~ 30 nm) line along the bottom-left of the image representing a fold in the suspended graphene sheet. Figure S3c is a color overlay of the inverse Fourier Transforms of the diffraction spots highlighted in Figure S3b, showing that the different grains meet along the lines of adsorbed contamination. We identified the contamination material decorating the grain boundaries using core-loss electron energy loss spectroscopy, which revealed that the contamination contains iron, oxygen, and carbon. The iron contamination is likely deposited during the ferric chloride etch used to remove the graphene from the copper substrate.

## 4. DF-TEM image overlay procedure

Figure S4a shows the raw DF-TEM data used to create the composite color DF-image shown in Figure 2g. We use GIMP 2 to do the overlay image processing (though any image processing software will do). First, each raw DF-image is read in as a layer and aligned by hand to the other layers if necessary. Next, the images are adjusted to maximize brightness and contrast, making sure to adjust each image to the same brightness/contrast levels. Each layer is then colorized according to the color code on the boxes, and the layers are merged. The levels in the final image are adjusted, clipping the highest and lowest intensities to enhance the image contrast. The overall color balance may be adjusted to enhance the color contrast in the image, giving the final image shown in Figure 2g. A similar process is used for all composite DF-images in Figure 2.

## 5. DF-TEM statistics acquisition

In order to extract the statistics shown in Figure 3, grain sizes and orientations were measured on three different samples.

To measure grain sizes, we determined the size of grains using raw DF-TEM images such as those shown in Figure S4a. The original image contrast was too low to be extracted by simple thresholding, but high enough that grains were clearly recognizable. To make size determination easier, we first traced the edge of each grain by hand using the Magnetic Lasso tool in Photoshop and then filled them with color. The images were then fed into ImageJ where the grains were picked out by thresholding and their areas were measured. With these methods, we counted 535 grains, shown in the histogram in Figure 3a.

To get relative grain orientations, we referred to DF-TEM composite images and their corresponding diffraction patterns. An example is shown in Figure S4d, where the measured angles are displayed over the grain boundaries in question. We recorded 238 data points on 8 different membranes. Error in this measurement varies by data point and is typically ±2º, though it may be up to ±5º, depending on whether it is clear which diffraction peak results in the grains shown. The upper bound on the angle is determined by the size of the objective aperture, and applies to highly polycrystalline regions with very closely spaced diffraction peaks.

## 6. More statistics on grain angles

For each image in Figure 3c, we averaged diffraction patterns sampled from 50 different membranes taken from a 1200 μm$^2$ region on a TEM grid. Each diffraction pattern is taken using a ~1μm diameter selected area aperture to exclude the SiN grid support. In Figure S5a, we reproduce Figure 3c(left) in the red box and also show data from an additional sample (blue box) In Figure S5b, we show the diffraction data as a polar intensity plot at 2.1 Å, with the trace color corresponding to the box color on the diffraction data in Figure S5a. In each diffraction pattern, there are small sub-peaks with ~ 5°-7° spacings, which are not as easily visible in the diffraction images.

Figure S5c shows a histogram of relative grain angle measured using STEM on 42 grain boundaries. This histogram shows the same peaks at low and high grain angles found in the grain angle histogram measured using DF-TEM.

## 7. Grain imaging comparisons: DF-TEM, STEM, AFM, SEM

Figure S6 demonstrates that decoration allows us to see the grain boundaries using a variety of microscopy techniques in addition to DF-TEM and ADF-STEM. Figures S6a-b show the same region of suspended graphene measured using DF-TEM and SEM. These images show a strong correlation between grain boundaries and contamination lines seen in SEM. Similarly, Figure

S6c-d show the same region of suspended graphene measured using STEM and AFM phase imaging. The decoration makes the grain boundaries visible because it has increased electron-sample interaction in SEM and STEM, and because it changes the tip-surface interaction in AFM. For these imaging techniques, the graphene needs to be suspended and relatively clean. Unfortunately, we find that doing photolithography on the graphene often deposits enough carbon and other surface particles to obscure the grain boundaries.

## 8. AFM indentation

To measure the mechanical properties of graphene by AFM indentation measurements, a procedure like the one described by Lee et al. (for exfoliated graphene) was employed[3]. We can use the model described in that study to measure the 2D elastic modulus, obtained by fitting deflection curves to the following equation:

(1) $$F = \sigma \pi d + E \frac{(qd)^3}{a^2}$$

Where $\sigma$ is the 2D pretension and $E$ is the 2D elastic modulus, $a$ is the radius of the graphene sheet, $d$ is the deflection of the graphene at its center, and $q$, a function of the Poisson's ratio, is taken to be 1.02. We found smaller values for the effective elastic modulus, a factor of ~6 smaller than those reported by Lee et al., but further discussion on the possible causes of this diminished elastic response lies outside the scope of this paper, and is still work in progress.

The force is calculated by the simple equation $F = kd_{tip}$ where $k$ is the spring constant of the cantilever and $d_{tip}$ is the tip deflection. Graphene's deflection is calculated by subtracting the tip deflection from the Z position of the AFM piezo (Z sensor). Plots of the Z sensor and deflection are shown in Figure S7.

The breaking load was read from the force curves as the point where the force exerted on the tip returns to zero or nearly zero. Smaller breaking events, where the force experienced only small drops, could be observed in some force plots, suggesting that smaller tears in graphene can occur before its complete failure.

As mentioned in the main text, we found that failure occurs at loads of ~100 nN on average. This is for graphene membranes 3.5 µm in diameter. In their study on exfoliated graphene[3], Lee et al. report a mean breaking force of ~1.7 µN, independent of membrane diameter (for diameters of 1.0 and 1.5 µm).

## 9. AC-EFM data

Figures S8c-d show the topography and phase images of an electrically contacted suspended graphene device, which correspond directly to the device schematic shown in Figure S8a. Unlike previously shown phase images, no grains are visible on the graphene surface because these features are obscured by extra contamination accumulated during the lithographic shaping. We performed AC-EFM measurements[4] on electrically contacted suspended graphene membranes using the circuit shown in S8b. Figure S8e shows the measured signal when driving the left electrode, the right electrode, and both electrodes respectively. By taking the ratio of the signals when the device is driven on one side and on both sides, we cancel out signals due to contamination and changing materials and measure the relative electrostatic potential along the device. Figure S8f shows the ratios of the data in Figure S8e with the images X-Y correlated to account for spatial drift, and rescaled to exclude external resistance. Figure S8g shows the relative potential from a single line scan from one electrode to another taken along the arrow shown in Figure S8f.